\documentstyle[aas2pp4]{article}

\slugcomment{Accepted by Ap. J. Letters}

\lefthead{M\'endez et al.}  \righthead{The speed of the LSR}

\begin{document}

\title{A large local rotational speed for the Galaxy found from
proper-motions: Implications for the mass of the Milky-Way}

\author{Ren\'e A. M\'endez}
\affil{Cerro Tololo Inter-American Observatory, Casilla 603, La
Serena, Chile. \\ E-mail: rmendez@noao.edu}

\author{Imants Platais\altaffilmark{1}, Terrence M. Girard, Vera
Kozhurina-Platais, William F. van~Altena}

\affil{Department of Astronomy, Yale University, P. O. Box 208101, New
Haven, CT 06520-8101. \\ E-mail: imants, vera, girard, vanalten
@astro.yale.edu}

\altaffiltext{1}{Visiting Astronomer, Cerro Tololo Inter-American
Observatory. CTIO is operated by AURA, Inc. under contract to the
National Science Foundation.}

\begin{abstract}

Predictions from a Galactic Structure and Kinematic model are compared
to the absolute proper-motions of about 30,000 randomly selected stars
with $9 < B_{\rm J} \le 19$ derived from the Southern Proper-Motion
Program (SPM) toward the South Galactic Pole. The absolute nature of
the SPM proper-motions allow us to measure not only the relative
motion of the Sun with respect to the local disk, but also, and most
importantly, the overall state of rotation of the local disk with
respect to galaxies. The SPM data are best fit by models having a
solar peculiar motion of +5~km~s$^{-1}$ in the V-component (pointing
in the direction of Galactic rotation), a large LSR speed of
270~km~s$^{-1}$, and a disk velocity ellipsoid that points towards the
Galactic center. We stress, however, that these results rest crucially
on the assumptions of both axisymmetry and equilibrium dynamics.

The absolute proper-motions in the U-component indicate a solar
peculiar motion of $11.0 \pm 1.5$~km~s$^{-1}$, with no need for a
local expansion or contraction term.

The implications of the large LSR speed are discussed in terms of
gravitational mass of the Galaxy inferred from the most recent and
accurate determination for the proper-motion of the LMC. We find that
our derived value for the LSR is consistent both with the mass of the
Galaxy inferred from the motion of the Clouds ($3 - 4 \times 10^{12}
M_\odot$ to $\sim 50$~kpc), as well as the timing argument, based on
the binary motion of M31 and the Milky Way, and Leo~I and the Milky
Way ($\ge 1.2 \times 10^{12} M_\odot$ to $\sim 200$~kpc).

\end{abstract}


\section{Introduction}

In this paper, we present the main results from an analysis of the
kinematic data obtained in the context of a massive {\it absolute}
proper-motion survey, the Southern Proper-Motion Program (SPM
hereafter). Our analysis is based on the SPM Catalog as described by
Platais et al. (1998). This catalog provides positions, absolute
proper-motions, and $BV$ photometry for about 30,000 randomly selected
stars, among other objects, near the South Galactic Pole. The sky
coverage of the SPM Catalog is about 720 deg$^2$ in the magnitude
range $5<V<18.5$. The accuracy of individual absolute proper-motions
is 3-8~mas~yr$^{-1}$ depending on the star's magnitude. For a complete
description of the catalogue structure, contents, plate measurement
and other astrometric/photometric details, the reader is referred to
Girard et al. (1998, Paper I) and Platais et al. (1998, Paper II). A
detailed analysis of the velocity distribution function, and star- and
color-counts for stars from the disk, thick-disk, and halo derived
from the SPM photometry and astrometry is addressed by M\'endez et
al. (1999, Paper III).

The arrangement of this Letter is the following: In Section~2 we
present an overview of the SPM photometric and proper-motion data. In
Section~3 the model used to compare the observed proper-motion
distributions with those predicted is described, while Section~4
presents the model predictions as compared to the proper-motion
data. Section~5 presents the implications of our results for the
global mass of the Galaxy. Finally, Section~6 presents the main
conclusions of the paper.

\section{The randomly selected sample} \label{sample}

The random SGP-SPM sample has been chosen such that, at a given ${\rm
B_J}$ magnitude interval, a fixed number of stars are randomly
extracted from all stars available in that magnitude interval. As a
result, one-magnitude intervals are chosen in the range $9 \le {\rm
B_J} \le 19$ for the subsequent analysis. At a given magnitude
interval, the complete sample and the randomly selected sample differ
only by a scale factor. Therefore, color and kinematic properties
binned in the proper magnitude intervals should be very similar to
those derived from a complete sample, except for the increased
uncertainties on the derived values because of the smaller sample - an
effect that is fully taken into account in the analysis below.

\section{The starcounts and kinematic model} \label{model} 

We model the starcounts concurrently with the kinematics by using the
Galactic structure and kinematic model presented by M\'endez and van
Altena (1996). The starcount model employed here has been tested under
many different circumstances, and has proved to be able to predict
starcounts that match the observed magnitude and color counts (in both
shape {\it and} number) to better than 10\%, and in many cases to
better than 1\%. The kinematic model presented by M\'endez and van
Altena (1996) has also been shown to be able to reproduce the
kinematics of disk stars toward different Galactic fields. In M\'endez
et al. (1999), a full analysis of the SPM-SGP data in terms of star
and color-counts, as well as the kinematics is presented. A basic
assumption in the M\'endez and van Altena (1996) model is that of the
Gaussian shape of the velocity ellipsoid. While the velocity
distribution observed for late-type stars (e.g., Dehnen 1998) is
clearly skewed and cannot be properly modeled by a Gaussian, the error
in the predicted median proper motion made by using a shifted Gaussian
is likely to be small. Indeed, M\'endez et al. (1999), show that
shifted-Gaussian velocity distribution functions do provide a good fit
to the full SPM proper-motion distributions as a function of apparent
magnitude, with perhaps a slight predicted excess of stars with
positive proper-motions in the direction of Galactic rotation.

Because of projection effects near the SGP, our model comparisons are
most sensitive to the motions in the direction of Galactic rotation
(the so-called V-component), and along the Galactic center-anticenter
direction (the U-component), and {\it not} to the motions
perpendicular to the Galactic plane (the W-component).

\subsection{The Solar peculiar velocity and the motion of the LSR}

There have been a number of determinations for the solar peculiar
motion. The classical result, quoted in Mihalas and Binney (1981),
gives $(U_\odot, V_\odot, W_\odot)= (+9.0, +12.0, +7.0)$~km~s$^{-1}$,
essentially based upon Delhaye's (1965) compilation. More recent
values for the solar peculiar motion do not seem to have converged to
a single value, especially in the V-component. Dehnen and Binney
(1998) find $(U_\odot, V_\odot, W_\odot)= (+10.00 \pm 0.36, +5.25 \pm
0.62, +7.17 \pm 0.38)$~km~s$^{-1}$ from a carefully selected unbiased
sample of Hipparcos stars, while Miyamoto and Zhu (1998) find
$(U_\odot, V_\odot, W_\odot)= (+10.62 \pm 0.49, +16.06 \pm 1.14, +8.60
\pm 1.02)$~km~s$^{-1}$ from 159 Cepheids, {\it also} from the
Hipparcos catalogue. As suggested earlier by Robin and Oblak (1987),
such variation in the $V_\odot$ component is probably due to
difficulties in separating the asymmetric drift from the intrinsic
solar motion, and also because of the peculiar motions exhibited by
the very young OB stars, which are still moving under the influence of
the spiral arm kinematics and/or of their parent molecular cloud.

In 1985, the IAU adopted a value for the motion of the LSR of
220~km~s$^{-1}$. Kerr and Lynden-Bell (1986) have discussed
extensively the determinations of $V(R_0)$, as well as $R_0$, and the
Oort constants A and B available until then. From a straight mean of
different determinations they obtained $V_{\rm LSR}(R_0)= 222 \pm
20$~km~s$^{-1}$ (their Table~4), while from independent determinations
of $R_0$, A, and B (their Tables~3 and~5), we find $V_{\rm LSR}(R_0)=
226 \pm 44$~km~s$^{-1}$. Although the uncertainties involved in
$V_{\rm LSR}(R_0)$ are larger than those of $V_\odot$, we shall see
that the relative motion of disk stars is more affected by
uncertainties in the Solar peculiar motion than uncertainties in the
motion of the LSR, since the whole nearby disk population is
approximately co-moving with the LSR.

\section{Model comparisons to the absolute proper-motions}
\label{modecoabprop}

The basic standard kinematic parameters employed in the model are
described in M\'endez and van Altena (1996) and M\'endez et
al. (1999). The standard model adopts $(U_\odot, V_\odot, W_\odot)=
(+11.0, +14.0, +7.5)$~km~s$^{-1}$ for the peculiar solar motion, and a
flat rotation curve with $V_{\rm LSR} (R_0)=+220$~km~s$^{-1}$. Because
we are analyzing data near the SGP, our model predictions do not span
a large range in Galactocentric distance, and therefore the model is
insensitive to the value adopted for the {\it slope} of the rotation
curve near $R_0$ (although see M\'endez et al. 1999). As for the disk
kinematics, we have adopted the velocity dispersions given in M\'endez
and van Altena (1996), with the scale-heights adopted in M\'endez et
al. (1999) which reproduce the observed SPM color- and magnitude
counts, a scale-length of 3.5~kpc (again, the model is not sensitive
to this last parameter, as it enters as a function of the
Galactocentric distance, which for the SPM-SGP data is
quasi-constant), and a value of $q=0$ for the velocity ellipsoid (see
Eq.~4 in M\'endez and van Altena 1996); $q=0$ implies a velocity
ellipsoid parallel to the Galactic plane at all heights from the
Galactic plane, while $q=1$ is for a velocity ellipsoid that points
towards the Galactic center). As shown by M\'endez et al. (1999) the
exact choice of kinematical parameters for the thick-disk and halo are
not critical to the conclusions of this Letter. The mean velocity of
stars with different velocity dispersion is computed via the proper
moments of the Collisionless Boltzmann equation (CBE). The CBE allow
us to relate our adopted velocity dispersions to the density laws, and
to derive an expression for the velocity lag as a function of the net
rotation of the disk (the local LSR) and the pressure support provided
by the velocity dispersions (see, e.g., Kuijken and Gilmore 1989,
Gilmore et al. 1989). We must emphasize however that, if the data
contain information that is inconsistent with the assumptions made (in
particular, the validity of an axisymmetric CBE, and the neglect of
stellar streams), then these assumptions inevitably could lead to
biased results -- the classical problem of all parametric methods.

\subsection{The Solar motion and the LSR speed}
\label{solmotlsr}

Figures~(\ref{fig1lett}) and~(\ref{fig2lett}) show the model-predicted
median absolute proper-motions for the standard model parameters
compared to the observations. These figures clearly show that, while
the standard model produces a very good fit to the U-component of the
proper-motion (along the Galactocentric direction), the V-component
(along Galactic rotation) is grossly underestimated, especially for
$B_{\rm J} < 14$, where the disk component dominates the overall
kinematics.

A radical change in the model predictions occurs when we change the
peculiar Solar motion from the standard value of $V_\odot =
+14$~km~s$^{-1}$ to $V_\odot = +5$~km~s$^{-1}$. The largest effect
occurs at the brightest bins, i.e., for nearby disk stars where the
Solar peculiar motion dominates the reflex motion. On the other hand,
we may keep the conventional Solar peculiar velocity, but change the
overall rotation speed for the LSR from the IAU adopted value of
+220~km~s$^{-1}$ to +270~km~s$^{-1}$, as suggested by recent Hipparcos
results (Miyamoto and Zhu 1998). In this case, the change in the
predicted motion becomes more important at fainter magnitudes where
one is sampling objects from the other Galactic components located at
larger distances, and where the dominant effect is that of the overall
rotation of the disk with respect to galaxies, and the relative state
of rotation between the different Galactic components. At the
brightest bins, the effect of changing $V_{\rm LSR} (R_0)$ becomes
also noticeable because of the larger fraction of bright giants, which
can be seen to large distances and, hence, the differential rotation
effects become amplified. A model in which we simultaneously change
the Solar peculiar motion {\it and} the LSR rotational speed to
+5~km~s$^{-1}$ and +270~km~s$^{-1}$, respectively, produces a much
better fit to the observed motions. We note that, in this model, the
velocity lags for the thick-disk and halo are increased in proportion
to the increase of the disk's rotational speed, as the net rotation of
those two components is kept constant at 180~km~s$^{-1}$ and
0~km~s$^{-1}$, respectively (for details see M\'endez et al. 1999). We
conclude that a model with $V_\odot = +5$~km~s$^{-1}$ and $V_{\rm LSR}
(R_0) = 270$~km~s$^{-1}$ clearly provides a much better fit to the
median motion in V than does the standard model with $V_\odot =
+11$~km~s$^{-1}$ and $V_{\rm LSR} (R_0) = 220$~km~s$^{-1}$.

The predicted median proper-motion in the U-component shows a good fit
to the observed values, and the changes in the V-component described
above do not affect this parameter substantially because of the
orthogonality of the projection effects toward the Galactic
poles. These results do show us, though, that the adopted value for
the Solar peculiar motion in this direction is the correct
one. Figure~(\ref{fig1lett}) shows the effect of changing the standard
$U_\odot = +11$~km~s$^{-1}$ by $\pm$3~km~s$^{-1}$. An eye-ball fit
from Figure~(\ref{fig1lett}) suggests for the U component of the solar
motion a value of $U_\odot = +11.0 \pm 1.5$~km~s$^{-1}$. Also, there
is no indication of a local expansion or contraction of the Galactic
disk, as is also found from the kinematics of local molecular clouds
(Belfort and Crovisier 1984).

A change of the orientation of the velocity ellipsoids from a
cylindrical to a spherical projection tends to produce a slightly
larger value for the velocity lag, especially at fainter
magnitudes. The upper solid line in Fig.~(\ref{fig2lett}) shows that,
while the fit of the model to the observed data for $V_\odot =
+5$~km~s$^{-1}$, $V_{\rm LSR} (R_0)=+270$~km~s$^{-1}$ is good at
bright magnitudes ($B_{\rm J} < 14$), the model underestimates the lag
at fainter magnitudes. However, by changing the orientation of the
velocity ellipsoid we can actually increase the lag. This is shown in
Figure~(\ref{fig2lett}) where a model with $V_\odot = +5$~km~s$^{-1}$,
$V_{\rm LSR} (R_0)=+270$~km~s$^{-1}$, and $q=1$ gives a better
overall fit to the observed median motion in the V-component than a
model with $q=0$.

\subsection{Implications for the mass of the Galaxy}\label{implimass}

The large LSR speed suggested by the SPM data poses an interesting
problem to dynamical models of the Galaxy. Since the LSR is, by
definition, in centrifugal equilibrium with the Galactic potential, a
larger LSR velocity implies a larger mass for the Galaxy interior to
the Solar circle. As emphasized by Rohlfs and Kreitschmann (1988),
practically all Galactic-mass models follow, to within 10\%, the
relation:

\begin{equation}
\frac{M(r)}{10^{10} M_\odot} = 2.3 \times 10^{-5} \, \left(
\frac{r}{\rm kpc} \right) \left (\frac{V_{\rm rot}(r)}{\rm {km \,
s^{-1}}} \right) ^2
\end{equation}

where $M(r)$ is the mass interior to the radius $r$, and $V_{\rm
rot}(r)$ is the circular speed at a distance $r$ from the Galactic
center. We see that an increase in the LSR speed from 220~km~s$^{-1}$
to 270~km~s$^{-1}$ would imply as much as 50\% more mass interior to
the Solar circle. Is there any other evidence that this might be
plausible?

An indication that this is perhaps not only possible but, indeed,
dynamically required, comes from a recent measurement of the absolute
proper-motions of the Large Magellanic Clouds (LMC) by Anguita et
al. (1999). They have used QSOs as an inertial reference frame to
obtain absolute proper-motions of stars in three fields toward the
LMC. The proper-motions are based on CCD frames taken at 7 to 12
epochs spanning eight years, with a mean accuracy for the
proper-motions from these three fields of $\pm
0.2$~mas~yr$^{-1}$. Their derived motion for the LMC is larger than
that inferred by previous photographic studies that have used either
bright stars or extended galaxies as a reference frame. Anguita et
al. derive a mass for the Milky Way out to 50~kpc of $3 - 4 \times
10^{12} M_\odot$, in contrast to previous determinations (e.g., Lin
and Lynden-Bell 1982) that implied a value of $7 \times 10^{11} \,
M_\odot$. Anguita's result is fully consistent with our findings:
Galactic mass models predict that the mass interior to the Solar
circle is only about 4\% of the total Galactic mass (Rohlfs and
Kreitschmann 1988). The range of mass values found by Anguita et
al. implies, through Eq.~(1) for $r=8.5$~kpc, a rotational velocity at
the Solar Galactocentric distance in the range $250-286$~km~s$^{-1}$,
which indeed brackets our results. Interestingly enough, a mass of $7
\times 10^{11} M_\odot$ implies a very improbable rotational speed of
120~km~s$^{-1}$.

On the other hand, our derived value for $V_{\rm LSR} (R_0)$ seems to
be in conflict with a recent determination of the proper motion of
Sgr~A$^\star$ by Reid et al. (1999), and Backer \& Sramek (1999) who
find $V_{\rm LSR} (R_0)/R_0\approx 29~{\rm km}\,{\rm s}^{-1}\,{\rm
kpc}^{-1}$ (employing the good assumption that Sgr~A$^\star$ is at
rest), which would require $R_0$ larger than 9~kpc, inconsistent with
almost all previous determinations. Could this inconsistency point,
perhaps, to evidence against our basic model assumptions which would
then lead to a biased $V_{\rm LSR} (R_0)$?

\subsection{Other evidence for a larger Galactic mass} \label{otherimpl}

Recently, Zaritsky (1999) has reviewed most of the evidence regarding
the mass of the Galactic halo and, by extension, the total mass of the
Galaxy. He points out that a squared ratio of rotational circular
velocities of 1.5 (such as that found from the 220 to 270~km~s$^{-1}$)
is not considered a serious discrepancy in the modeling of the total
Galactic mass. He furthermore points out that timing arguments
involving the relative motion of M~31 and the Milky~Way imply a mass
for the latter of $\sim \, 1.4 \times 10^{12} \, M_\odot$. A similar
argument, based on the Leo~I-Milky-Way system, leads to a mass of
$1.1-1.5 \times 10^{12} \, M_\odot$. In both cases, the analysis
excludes angular momentum, overlapping mass distributions at earlier
times, and the evolution (growth) of the Galactic halo with age, all
of which would lead to an increased total mass. Zaritsky also reviews
an extension of the two-body calculations into the larger environment,
which leads to an even larger mass, in the range $1.9 - 2.3 \times
10^{12} \, M_\odot$, concluding that all this evidence implies a mass
out to $\sim \, 200$~kpc of $\ge 1.2 \, \times 10^{12} \,
M_\odot$. Thus, Anguita's results based on the LMC proper motions, and
its correspondence to our findings, are in line with Zaritsky's
arguments.

\section{Conclusions}

We have used the distribution of absolute proper motions of $\sim
30,000$ randomly selected stars, part of the Southern Proper Motion
survey, to constrain a structural \& kinematic model of the
Galaxy. The absolute proper-motions in the U-component indicate a
solar peculiar motion of $11.0 \pm 1.5$~km~s$^{-1}$, with no need for
a local expansion or contraction term. In the V-component, the
absolute proper-motions can only be satisfactorily reproduced by the
model if we adopt a solar peculiar motion of +5~km~s$^{-1}$, a large
LSR speed of 270~km~s$^{-1}$, and a (disk) velocity ellipsoid that
always points towards the Galactic center.

The larger than expected LSR speed leads to a larger mass for the
Galaxy interior to the Solar Circle by a factor of about 1.5. This is
compared with other recent evidence indicating that the Milky-Way
might indeed be more massive than previously thought. This evidence
comes mainly from a new measurement of the proper motion of the LMC
(Anguita et al. 1999) and from the binary motion (``timing argument'')
of the M31 - Milky Way and Leo~I - Milky Way pairs (Zaritsky
1999). Our larger LSR speed is also coincident with a similar finding
by Miyamoto and Zhu (1998) using O-B5 stars and Cepheids from the
Hipparcos catalogue. We stress that our result of a large LSR speed
rests crucially on the assumptions of both axisymmetry and equilibrium
dynamics. Failure to meet these conditions would lead to
inconsistencies, such as those related to the Solar-galactocentric
distance indicated at the end of Section~4.2.

High accuracy proper motion measurements for the SMC, and proper
motions for some of the nearby dwarf Spheroidals will help in
confirming a larger value for the mass of the Galaxy, while future
space interferometric space missions like FAME ({\tt
http://www.usno.navy.mil/fame}), GAIA ({\tt
http://astro.estec.esa.nl/GAIA/}), and SIM ({\tt
http://sim.jpl.nasa.gov/}), will directly measure the disk rotational
speed across the entire Galaxy.

\acknowledgments

The SPM is a joint project of the Universidad Nacional de San Juan,
Argentina and the Yale Southern Observatory. Financial support for the
SPM has been provided by the US NSF and the UNSJ through its
Observatorio Astron\'omico "F\'elix Aguilar". We would like to also
acknowledge the invaluable assistance of Lic. Carlos E. L\'opez, who
participated in, and supervised, all of the SPM second-epoch
observations. We are indebted to an anonymous referee who provided
many useful comments on the limitations and potentialities of the
Galactic model employed here. These, and other comments by the
referee, have greatly helped to clarify many points of the original
manuscript.

\clearpage

\figcaption[fig1lett.ps]{Median observed (filled dots with error bars)
and predicted (lines) absolute proper-motions along U (Galactocentric
direction). In some cases the error bars are smaller than the size of
the dot. The solid line is for the standard model, the dashed line is
for a model with a Solar peculiar motion in the U-component of
+8~km~s$^{-1}$, while the dot-dashed line is for a model with $U_\odot
= +14$~km~s$^{-1}$. The standard model, with $U_\odot =
+11$~km~s$^{-1}$ provides the best fit to the U-component of the
observed motion. The agreement to the model also indicates the absence
of any significant expansion/contraction in the local disk (see text).
\label{fig1lett} }

\figcaption[fig2lett.ps]{Median observed (filled dots with error bars)
and predicted (lines) absolute proper-motions along V (Galactic
rotation). In some cases the error bars are smaller than the size of
the dot. The lower solid line is from the standard model with $V_\odot
= +14$~km~s$^{-1}$, and $V_{\rm LSR} (R_0)=+220$~km~s$^{-1}$. The
dashed line is for a model with a disk having a rotational speed of
+270~km~s$^{-1}$. The dotted line indicates the predictions for a
model with a Solar peculiar motion in the V-component of
+5~km~s$^{-1}$ instead of the classical value +14~km~s$^{-1}$ adopted
in the standard model, while the triple-dot dash line shows a model
with, both, $V_\odot = +5$~km~s$^{-1}$, and $V_{\rm LSR}
(R_0)=+270$~km~s$^{-1}$. Clearly, a larger LSR speed improves the
model predictions at the fainter ($B_{\rm J} \ge 15$) bins. Finally,
the upper solid line is for a model with $V_\odot = +5$~km~s$^{-1}$,
$V_{\rm LSR} (R_0)=+270$~km~s$^{-1}$, and $q=1$, indicating that
the best fit to the observed median motion in the V-component is
provided by a velocity ellipsoid whose major axis points to the
Galactic center as we move away from the Galactic plane.
\label{fig2lett} }


\clearpage

\begin{thebibliography}{99}

\bibitem[Anguita et al. 1999]{anet99} Anguita, C., Loyola, P.,
Pedreros, M.H., 1999, Submitted to \aj

\bibitem[Backer and Sramek 1999]{basr99} Backer, D.C. \& Sramek, R.A.,
1999, astro-ph/9906048

\bibitem[Belfort, and Crovisier 1984]{bc84} Belfort, P., and
Crovisier, J., 1984, \aap, 136, 368

\bibitem[Delhaye 1965]{de65} Delhaye, J., 1965, in Galactic Structure,
Stars and Stellar Systems, Vol. 5, eds.  A. Blaauw and
M. Schmidt (The University of Chicago, Chicago), page 61

\bibitem[Dehnen and Binney 1998]{debi98} Dehnen, W., Binney, J. J.,
1998, \mnras, 298, 387

\bibitem[Dehnen 1998]{de98} Dehnen, W., 1998, \aj, 115, 2384

\bibitem[Gilmore et al. 1989]{gwk89} Gilmore, G., Wyse, R. F. G., and
Kuijken, K., 1989, \araa, 27, 555

\bibitem[Girard et al. 1998]{gir98} Girard, T.M., Platais, I.,
Kozhurina-Platais, V., van Altena, W.F., \& L\'{o}pez, C.E. 1998, \aj,
115, 855 (Paper I)

\bibitem[Kerr and Lynden-Bell 1986]{kelb86} Kerr, F. J., and
Lynden-Bell, D., 1986, \mnras, 221, 1023

\bibitem[Kuijken and Gilmore 1989]{kg89} Kuijken, K., and Gilmore, G.,
1989, \mnras, 239, 571

\bibitem[Lin and Lynden-Bell 1988]{lily88} Lin, D.N.C., and
Lynden-Bell, D., 1982, \mnras, 198, 707

\bibitem[M\'endez \& van Altena 1996]{men96} M\'endez, R.A., \& van
Altena, W.F. 1996, \aj, 112, 655

\bibitem[M\'endez et al. 1999]{men99} M\'endez, R.A., Platais, I.,
Girard, T.M., Kozhurina-Platais, V., van Altena, W.F. 1999, \aj,
submitted (Paper III)

\bibitem[Mihalas and Binney 1981]{mb81} Mihalas, D., and Binney, J.,
1981, Galactic Astronomy, San Francisco: Freeman

\bibitem[Miyamoto and Zhu 1998]{mizh98} Miyamoto, M., and Zhu, Z.,
1998, \aj, 115, 1483

\bibitem[Platais et al. 1998]{plet98} Platais, I., Girard, T.M.,
Kozhurina-Platais, V., van Altena, W. F., L\'opez, C. E., M\'endez,
R. A., Ma W.-Z., Yang, T.-G., MacGillivray, H. T., Yentis, D. J.,
1998, \aj, 116, 2556 (Paper II)

\bibitem[Reid et al. 1999]{reet99} Reid, M.J., Readhead, A.C.S.,
Vermeulen, R.C., Treuhaft, R.N., 1999, astro-ph/990575

\bibitem[Robin and Oblak 1987]{roob87} Robin, A. C., and Oblak, E.,
1987, Evolution of Galaxies, 10th IAU European Regional Meeting,
Vol. 4, J.  Palous, Ondrejov: Astronomical Institute of the
Czechoslovak Academy of Sciences, 323

\bibitem[Rohlfs and Kreitschmann 1988]{rokr88} Rohlfs, K., and
Kreitschman, J., 1988, \aap, 201, 51

\bibitem[Zaritsty 199]{za99} Zaritsky, D. 1999, in The Third Stromlo
Symposium: The Galactic halo, ASP Conf. Ser., Vol. 165,
eds. B. K. Gibson, T. S. Axelrod and M.E. Putnam (Astron. Soc. of the
Pacific, California), page 34

\end{thebibliography}
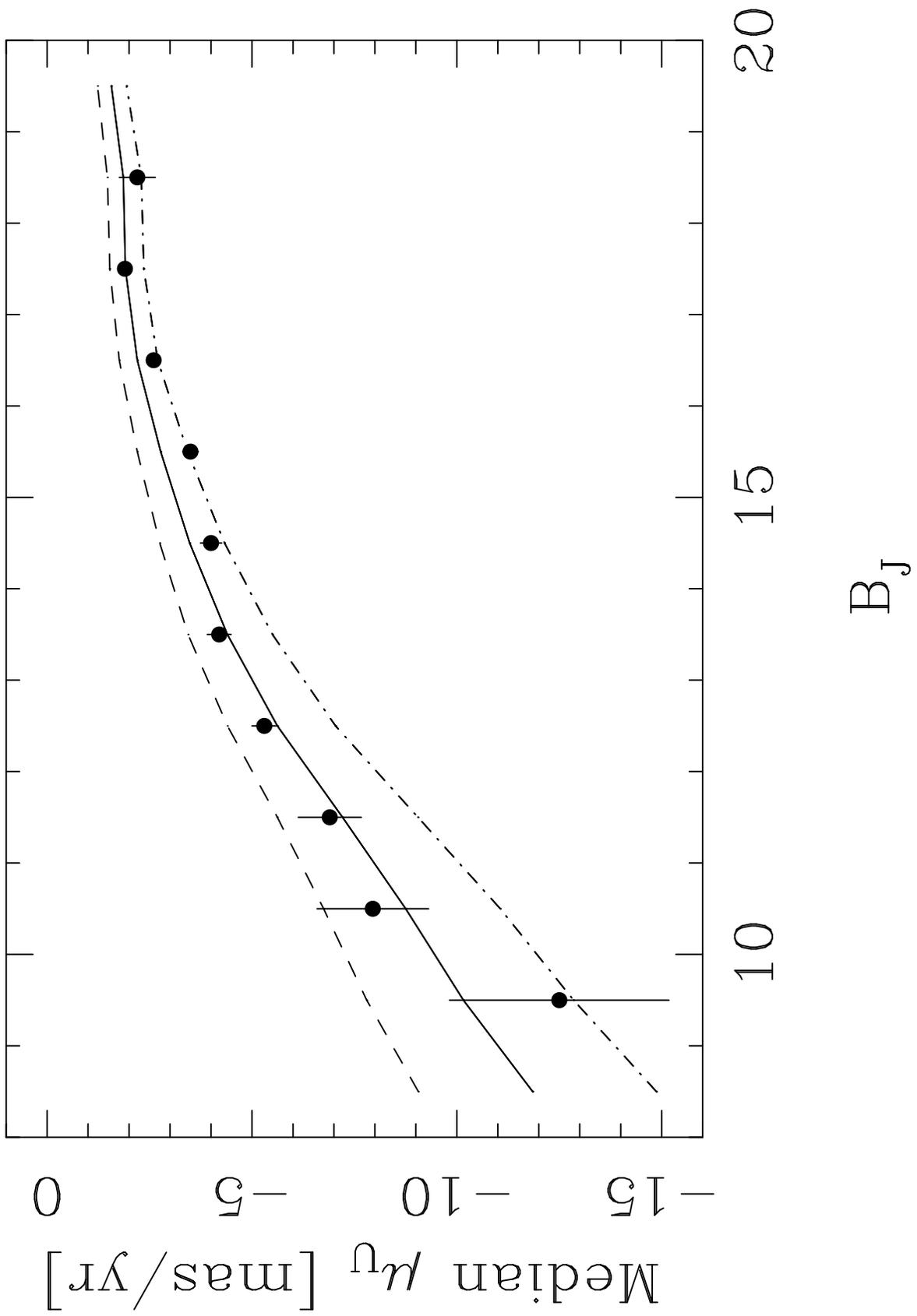
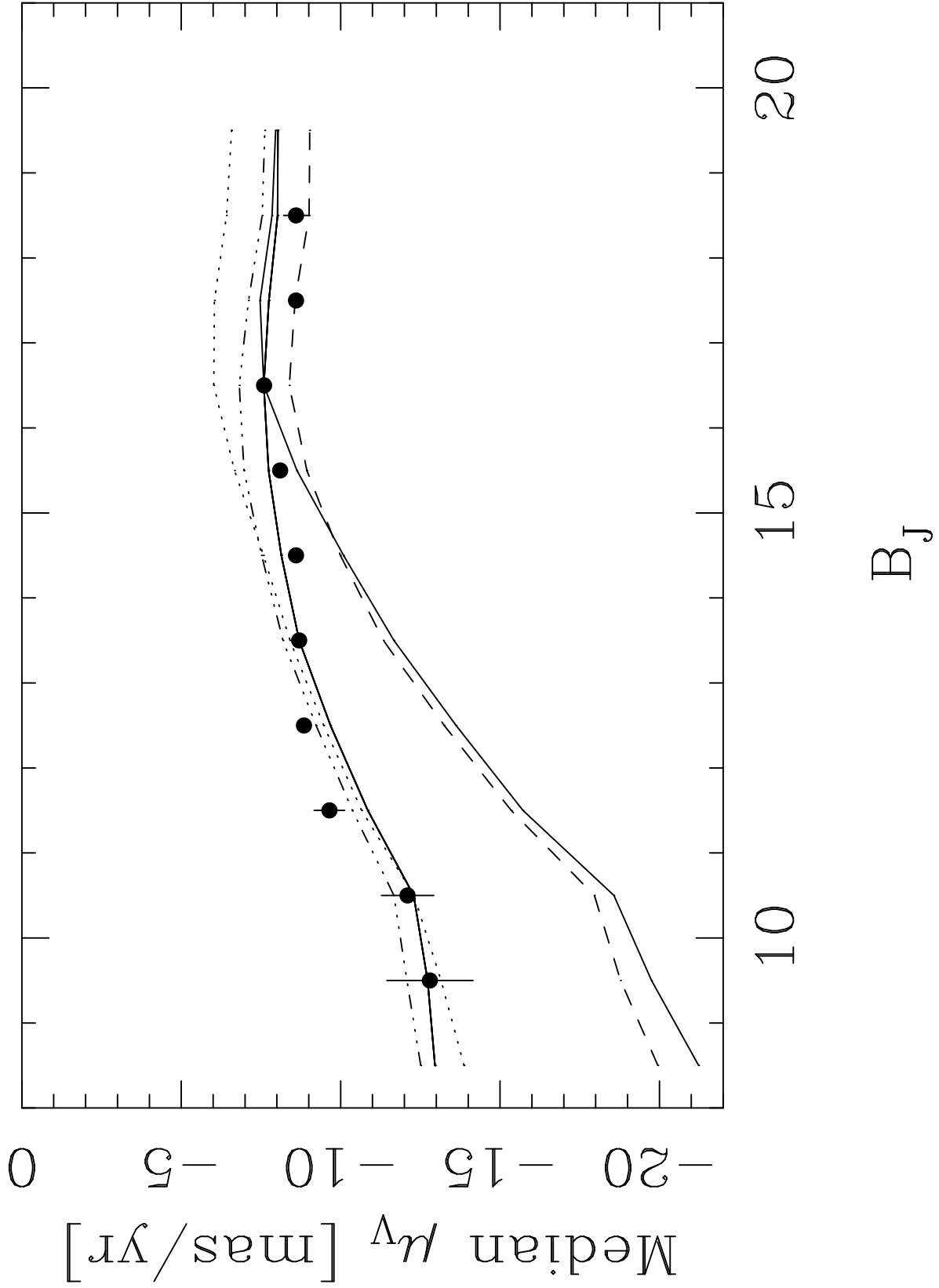
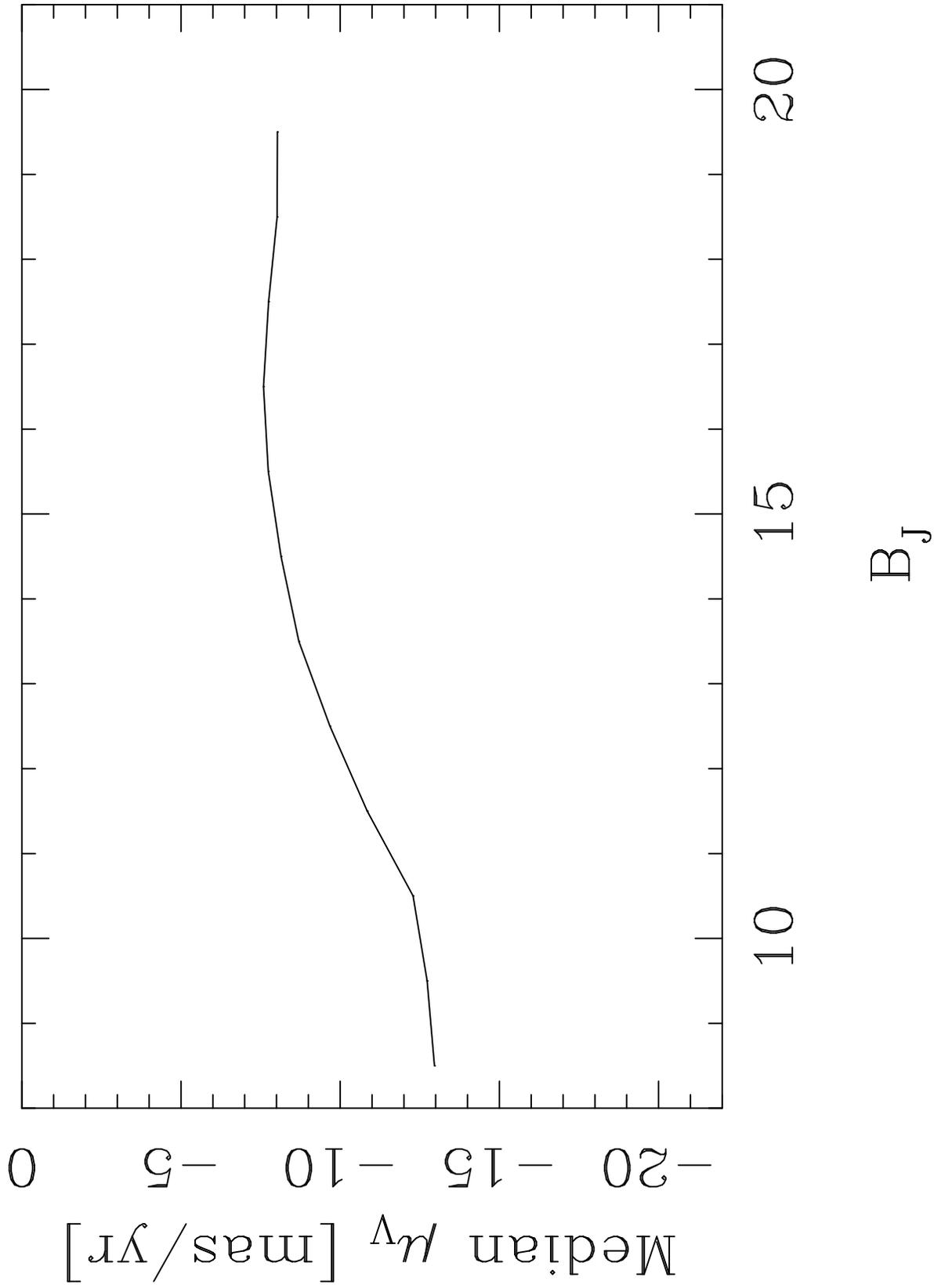
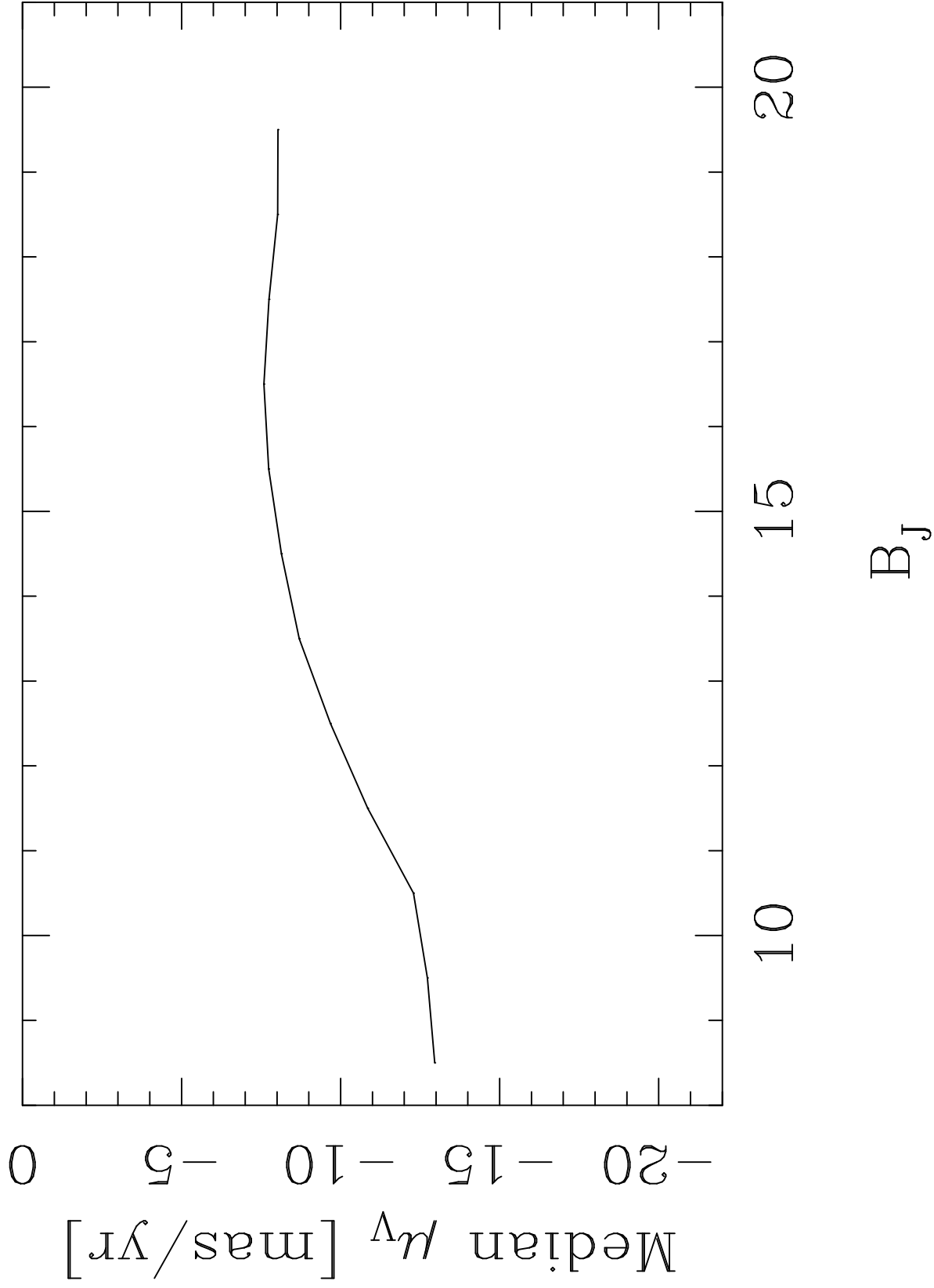
\end{document}